\documentclass[aps,pra,onecolumn,superscriptaddress,11pt]{revtex4}

\usepackage{graphicx}
\usepackage[usenames]{color}
\usepackage{epsfig}
\usepackage{verbatim}
\usepackage{listings}
\usepackage{rotating}
\usepackage{amsmath}
\usepackage{amssymb}
\usepackage{amsfonts}
\usepackage{dcolumn}
\usepackage{fancyhdr}
\usepackage{threeparttable}

\newcommand{\rpC}{\ensuremath{r_{\mathrm E}}}

\newcommand{\rZ}{\ensuremath{r_{\mathrm Z}}}

\newcommand{\rZt}{\ensuremath{R^3_{(2)}}}

\newcommand{\mup}{\ensuremath{\mu \mathrm{p} }}

\newcommand{\rr}{\ensuremath{\langle r^2 \rangle}}
\newcommand{\rrr}{\ensuremath{\langle r^2 \rangle^{3/2}}}

\newcommand\cnt[2]{\multicolumn{#1}{c}{#2}}
\newcommand\lft[2]{\multicolumn{#1}{c}{#2}}
\newcommand\ttxt[1]{\multicolumn{1}{r}{#1}}

\renewcommand{\baselinestretch}{1.2}

\newcolumntype{f}[1]{D{.}{.}{#1}}

\begin{document}


\title{Theory of the 2S-2P Lamb shift and  2S hyperfine splitting  in muonic hydrogen}

\author{Aldo Antognini}
\email{aldo@phys.ethz.ch}
\affiliation{Institute for Particle Physics, ETH Zurich, 8093 Zurich, Switzerland.}
\author{Franz~Kottmann}
\affiliation{Institute for Particle Physics, ETH Zurich, 8093 Zurich, Switzerland.}
\author{Fran\c{c}ois~Biraben}
\affiliation{Laboratoire Kastler Brossel, \'Ecole Normale Sup\'erieure, CNRS and 
             Universit\'e P.~et M.~Curie, 75252 Paris, CEDEX 05, France.}
\author{Paul~Indelicato}
\affiliation{Laboratoire Kastler Brossel, \'Ecole Normale Sup\'erieure, CNRS and 
             Universit\'e P.~et M.~Curie, 75252 Paris, CEDEX 05, France.}
\author{Fran\c{c}ois~Nez}
\affiliation{Laboratoire Kastler Brossel, \'Ecole Normale Sup\'erieure, CNRS and 
             Universit\'e P.~et M.~Curie, 75252 Paris, CEDEX 05, France.}
\author{Randolf~Pohl}                        
\affiliation{Max--Planck--Institut f{\"u}r Quantenoptik, 85748 Garching, Germany.}
%

\begin{abstract}
The 7$\sigma$ discrepancy between the proton rms charge radius from
muonic hydrogen and the CODATA-2010 value from hydrogen spectroscopy and
electron-scattering has caused considerable discussions. Here, we
review the theory of the 2S-2P Lamb shift and 2S hyperfine splitting
in muonic hydrogen combining the published contributions and
theoretical approaches.  The prediction of these quantities is
necessary for the determination of both proton charge and Zemach radii
from the two 2S-2P transition frequencies measured in muonic
hydrogen~\cite{Pohl:2010:Nature_mup1,Antognini:2012:Nature_mup2}.
%
  %
  %
%
\end{abstract}
\noindent

\maketitle

\section{Introduction}

The study of energy levels in 
hydrogen~\cite{Parthey:2011:PRL_H1S2S, Beauvoir:2000:Hydeurydls},
hydrogen-like atoms like muonium~\cite{Chu:1998:Muonium1S2S}
or positronium~\cite{Danzmann:1989}, 
as well as free~\cite{Hanneke:2008} and bound~\cite{Sturm:Wagner:Blaum:2012}
electron g-factors provide the most accurate and precise verifications 
of quantum electrodynamics (QED)~\cite{Karshenboim:2005:PPS}.

Comparison of the measured transition frequencies in hydrogen with
theory is limited by the uncertainty of the proton
structure~\cite{Jentschura:Mohr:fundamental:constants:2008, Karshenboim:2005:PPS}.
Here, the main uncertainty originates from the root mean square (rms) charge
radius of the proton, defined as
$\rpC^2=\int \mathrm{d}^3r\, r^2 \rho(\mbox{\boldmath{$r$}})$, where
$\rho$ is the normalized charge density of the proton.
Muonium and positronium are made from point-like elementary particles and 
therefore do not suffer from uncertainties due to finite-size effects.
The ultimate experimental precision is however limited by the short
lifetime of these systems.
A way out to improve the test of hydrogen energy levels is given by
spectroscopy of muonic hydrogen (\mup{}), an atom formed by a muon and
a proton. This provides a precise determination of \rpC{}.

Recently the comparison of the measured
$2\mathrm{P}_{3/2}^{F=2}-2\mathrm{S}_{1/2}^{F=1}$ transition in muonic
hydrogen~\cite{Pohl:2010:Nature_mup1}
\begin{equation}
\label{eq:first-transition-exp}
\begin{array}{ll}
\Delta E_{2\mathrm{P}_{3/2}^{F=2}-2\mathrm{S}_{1/2}^{F=1}}^\mathrm{exp} &
 =  206.2949(32) ~ \textrm{meV}\\
\end{array}
\end{equation}
with the theoretical prediction based on bound-state
QED~\cite{Pachucki:1996:LSmup,Pachucki:1999:ProtonMup,
  Borie:2005:LSmup}, as summarized in the Supplementary Information
of~\cite{Pohl:2010:Nature_mup1}, 
from now on referred to as Ref.~\cite{Pohl:2010:Nature_mup1_suppl}

\begin{equation}
\label{eq:first-transition-th}
\begin{array}{ll}
\Delta E_{2\mathrm{P}_{3/2}^{F=2}-2\mathrm{S}_{1/2}^{F=1}}^\mathrm{th} &
  =  209.9779(49)-5.2262\, \rpC^2 +0.0347\, \rpC^3 \quad \textrm{meV}\\
\end{array}
\end{equation}
yielded $\rpC= 0.84184(67)$~fm ~\cite{Pohl:2010:Nature_mup1}.
Throughout the paper we assume radii to be in fm, resulting energies in meV.

The muonic hydrogen value of $\rpC$
is an order of magnitude more precise than the CODATA-2010 value
$\rpC^\mathrm{CODATA}=0.8775(51)$~fm~\cite{Mohr:2012:CODATA10} which 
originates from a least-square adjustment of measurements from hydrogen
spectroscopy\cite{Lundeen:1981:LS,Hagley:1994:FShyd,
Schwob:1999:Hydr2S12D,Beauvoir:2000:Hydeurydls,Fischer:2004:DriftFundConst,
Parthey:2010:PRL_IsoShift,Arnoult:Nez:1S3S:2010,Parthey:2011:PRL_H1S2S}
and electron-proton scattering~\cite{Sick:2003:RP,Blunden:2005:RP,
Bernauer:2010:NewMainz}.  However, the muonic hydrogen 
value is 4\% smaller than the CODATA value~\cite{Mohr:2012:CODATA10}.
This 7 standard deviations discrepancy is now commonly referred to as 
the ``proton radius puzzle''~\cite{Pohl:2011:ConfPSAS10}.

Many efforts have been made to solve the radius puzzle.
Theory of muonic hydrogen energy levels has been refined significantly,
and is reviewed here.
The suggestion~\cite{Jentschura:2011:AnnPhys2} that molecular effects due to
$p \mu e^-$ ion formation may be responsible for the discrepancy has been
ruled out recently~\cite{Karr:2012:3body}.
Electron scattering has seen a huge activity recently, both from new
measurements~\cite{Bernauer:2010:NewMainz, Distler:2011:Zemach,
  Zhan:2011:JLab_Rp} and new re-analyses of the world
data~\cite{CloetMiller:2011:3rdZemach, HillPaz:2010:Extrapolation, Ron:2011,
  Hasell:2011:BLAST, Sick:2011:Troubles, Gilad:2011, Adamuscin:2011,
  Adamuscin:2012, Sick:2012:Problems, Mueller:2012:MAMI}.
Several authors studied proton structure at low 
energies~\cite{
Pineda:2005:Lamb-shift:proton-radius-def,
Vanderhaeghen:2010,
Miller:2011:NaturalResolution,
HillPaz:2011:TwoPhotons,
Hagler:2011,
Collins:2011,
BirseMcGovern:2012,
LorenzHammerMeissner:2012, 
Ledwig:2012}.
Also, physics beyond the standard model has been 
considered~\cite{Jentschura:2011:AnnPhys2, Brax:2011, Rivas:2011,
  Karshenboim:2010:NewPhysics,Jaeckel:2010, Barger:2011,
  Tucker-Smith:2011, Batell:2011:PV_muonic_forces, Barger:2012,
  McKeen:Pospelov:2012, CarlsonRislow:2012:NewPhysisc}.
On the experimental side, several projects aim at a new determination of the
Rydberg constant ($R_\infty$) at MPQ, LKB~\cite{Arnoult:Nez:1S3S:2010},
NPL~\cite{Flowers:NPL:2007} and NIST~\cite{Tan:NIST:2011}, to rule out
possible systematics shifts in the previous measurements that determine
$R_\infty$~\cite{Biraben:2009:SpectrAtHyd}. A new measurement of the classical 
2S-2P Lamb shift in hydrogen is being prepared~\cite{Hessels:2012:PC}.
New electron scattering measurements are proposed or underway~\cite{Kohl:2009:OLYMPUS,DarkLightProposal:2010,Gasparian:2011:Proposal,Raue:2010:CLAS_TPE,Slifer:2012:NucleonSpin}. And finally, new measurements of muonic hydrogem deuterium or helium ions may in futire shed new light on the proton radius puzzle~\cite{Adamczak:2012:mupHFS,Antognini:2012:Conf:PSAS}.
%
%
Several studies have been concerned with the theory
of the $n=2$ energy levels in muonic hydrogen.
Here we summarize the various contributions of
these investigations updating the theoretical prediction of the muonic
hydrogen 2S-2P Lamb shift and 2S-HFS.
We anticipate already here that no big error or additional
contribution have been found which could solve the observed
discrepancy between experiment and theory of the 2S-2P energy
difference:
\begin{equation}
\label{eq:discrepancy}
\mbox{discrepancy} =\Delta E_{2\mathrm{P}_{3/2}^{F=2}-2\mathrm{S}_{1/2}^{F=1}}^\mathrm{exp}- \Delta E_{2\mathrm{P}_{3/2}^{F=2}-2\mathrm{S}_{1/2}^{F=1}}^\mathrm{th}(\rpC^\mathrm{CODATA})   
    =  0.31\,\textrm{meV}.
\end{equation}
This corresponds to a relative discrepancy of 0.15\,\%.
%
%

Two transition frequencies in muonic hydrogen have been
measured. One starts from the 2S-triplet state
$\nu_t=2\mathrm{P}_{3/2}^{F=2}-2\mathrm{S}_{1/2}^{F=1}$~\cite{Pohl:2010:Nature_mup1,
  Antognini:2012:Nature_mup2}, and the other from the 2S-singlet state
$\nu_s=2\mathrm{P}_{3/2}^{F=1}-
2\mathrm{S}_{1/2}^{F=0}$~\cite{Antognini:2012:Nature_mup2}, as shown in
Fig.~\ref{fig:energy-levels}.
As detailed in Sec.~\ref{sec:Lamb+HFS} we can deduce from these
measurements both the "pure'' Lamb shift ($\Delta E_\mathrm{L}=\Delta
E_\mathrm{2P_{1/2}-2S_{1/2}}$) and the 2S-HFS splitting ($\Delta
E_\mathrm{HFS}$), each independent of the other.
Comparing the experimentally determined $\Delta E_\mathrm{L}$ with its
theoretical prediction given in Sec.~\ref{sec:lamb-shift_rp} yields an
improved \rpC{} value free of uncertainty from the 2S-HFS.
Similarly, comparing the experimentally determined $\Delta
E_\mathrm{HFS}$ with its theoretical prediction given in
Sec.~\ref{sec:HFS_zemach} results in the determination of the Zemach
radius \rZ{}.

Perturbation theory is used to calculate the various corrections to
the energy levels involving an expansion of both operators and wave
functions.
The radiative (QED) corrections are obtained in an expansion in $\alpha$,
binding effects and relativistic effects in $(Z\alpha)$, and recoil
corrections in the ratio of the masses of the two-body system
$(m/M)$. $Z=1$ is the atomic charge number and $\alpha$ the fine
structure constant.  The contributions related to the proton
structure are in part described by an expansion in powers of \rpC{}
and \rZ{}.
The book-keeping of all the corrections contributing to the
\mup{} Lamb shift and 2S-HFS is challenging because:
\begin{itemize}
\item All corrections are mixed as $\alpha^x \, (Z \alpha)^y \; (m/M)^z \; \rpC^t$\,.
\item There are large finite-size and recoil ($m/M \approx 1/9$) corrections.
\item One cannot develop the calculation in a systematic way like in $g-2$ for free particles.
\item Widely different scales are involved: the masses, the three-momenta and the kinetic energies of the constituents.
\item Different authors use different terminologies for identical terms.
\item Different methods are being used: Schr\"odinger equation + Breit
  corrections versus Dirac equation, Grotch- versus Breit-type recoil
  corrections, all-order versus perturbative in $(Z\alpha)$ and finite-size,
  non-relativistic QED (NRQED) etc.
\end{itemize}

In this study we summarize all known terms included in the Lamb shift
and the 2S-HFS predictions which are used
in~\cite{Antognini:2012:Nature_mup2} to determine the proton charge
radius and the Zemach radius. The majority of these terms can be found
in the works of
Pachucki~\cite{Pachucki:1996:LSmup,Pachucki:1999:ProtonMup},
Borie~\cite{Borie:2005:LSmup}, and
Martynenko~\cite{Martynenko:2005:mupHFS2S,Martynenko:2008:HFS_Pstates_mup}.
These earlier works have been reviewed in Eides et
al.\cite{Eides:2001:HydTheo,Eides:2006:Book}.
After the publication of ~\cite{Pohl:2010:Nature_mup1}, a number of authors
have revisited the theory in muonic hydrogen, e.g.\
Jentschura~\cite{Jentschura:2011:AnnPhys1,Jentschura:2011:relrecoil},
Karshenboim {\it et al.}~\cite{Karshenboim:2012:relrecoil} and
Borie~\cite{Borie:2012:LS_revisited}.
Note that the arXiv version 1103.1772v6 of Borie's
article~\cite{Borie:2012:LS_revisited} contains corrections to the published
version, which is why we refer to ``Borie-v6'' here.
In addition Indelicato~\cite{Indelicato:2012} checked and improved
many of the relevant terms by performing numerical integration of the
Dirac equation with finite-size Coulomb and Uehling potentials.
Carroll {\it et al.} started a similar effort~\cite{Carroll:2011:NP}.

\section{The experimental Lamb shift and 2S-HFS}
\label{sec:Lamb+HFS}

\begin{figure}
\includegraphics[height=0.6\columnwidth, angle=0]{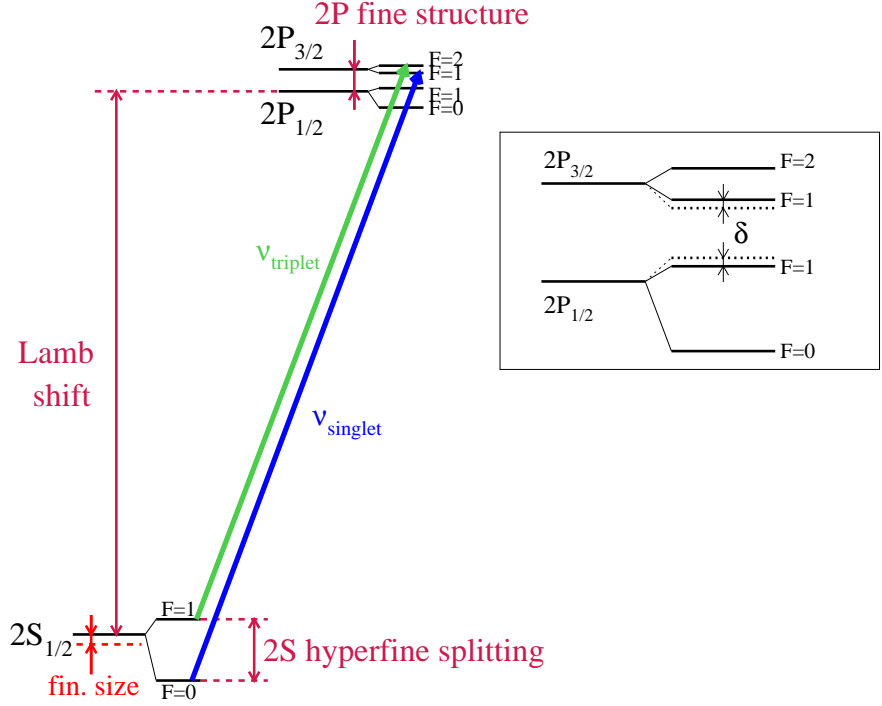}
\caption{2S and 2P energy levels. The measured transitions $\nu_t$
  \cite{Pohl:2010:Nature_mup1} and $\nu_s$
  \cite{Antognini:2012:Nature_mup2} are indicated together with Lamb
  shift, fine and hyperfine splittings and finite-size effects.  The
  main figure is drawn to scale.  The insets zooms in on the 2P
  states. Here, the mixing of the 2P(F=1) levels shifts them by $\pm
  \delta$ (see Eq.(\ref{eq:delta})).}
       \label{fig:energy-levels}
\end{figure}

The 2S and 2P energy levels in muonic hydrogen are presented in
Fig.~\ref{fig:energy-levels}. 
Because the measurements of the Lamb shift involve only $n=2$ states, 
the main term of the binding energy (632~eV given by the Bohr structure)
drops out and the results do not depend on the Rydberg constant.

The 2S-2P splitting arises from  relativistic, hyperfine, radiative, 
recoil, and nuclear structure effects.
The Lamb shift is dominated by the one-loop electron-positron vacuum
polarization of 205~meV.
The experimental uncertainty of the Lamb shift is of relative order
$u_r \approx 10^{-5}$.
Thus the various contributions to the Lamb shift should be calculated to
better than $\sim 0.001$~meV to be able to exploit the full experimental
accuracy.

The two measured transition frequencies shown in
Fig.~\ref{fig:energy-levels} are given by
\begin{equation}
\begin{split}
 h\nu_t =& E({\textstyle 2\mathrm{P}_{3/2}^{F=2}}) - E({\textstyle 2\mathrm{S}_{1/2}^{F=1}}) \\
        =& \Delta E_\mathrm{L} + \Delta E_\mathrm{FS} 
                + \frac{3}{8}\Delta E_\mathrm{HFS}^{2\mathrm{P}_{3/2}} 
                - \frac{1}{4}\Delta E_\mathrm{HFS} ,  \\[5mm]
 h\nu_s =& E({\textstyle 2\mathrm{P}_{3/2}^{F=1}}) - E({\textstyle 2\mathrm{S}_{1/2}^{F=0}}) \\
        =& \Delta E_\mathrm{L} + \Delta E_\mathrm{FS} 
                - \frac{5}{8}\Delta E_\mathrm{HFS}^{2\mathrm{P}_{3/2}} + \delta
                + \frac{3}{4}\Delta E_\mathrm{HFS} , 
\end{split}
\end{equation}
where $h$ is the Planck constant, and $\Delta E_\mathrm{FS}=\Delta
E_\mathrm{2P_{3/2}}-\Delta E_\mathrm{2P_{1/2}}$ is the fine structure splitting of
the 2P-state.
For the 2P fine structure splitting we use the value~\cite{Martynenko:2008:HFS_Pstates_mup}
\begin{equation}
\Delta E_\mathrm{FS} = 8.352082 ~ \textrm{meV},
\label{eq:fs_2S}
\end{equation}
which is in agreement with the values
8.3521\,meV~\cite{Borie:2012:LS_revisited} and
$8.351988-0.000052\,\rpC^2=8.351944$\,meV~\cite{Indelicato:2012}, Eq.~(120).
The $2\mathrm{P}_{3/2}$ hyperfine structure splitting
is~\cite{Martynenko:2008:HFS_Pstates_mup}
\begin{equation}
\Delta E_\mathrm{HFS}^{2\mathrm{P}_{3/2}} = 3.392588 ~ \textrm{meV},
\label{eq:hfs_2P}
\end{equation}
in agreement with 3.392511\,meV~\cite{Indelicato:Mohr:2012}.
The $2\mathrm{P}_{3/2}^{F=1}$ level is shifted upwards due to state
mixing~\cite{Romanov:1995} of the two 2P(F=1) levels
by~\cite{Martynenko:2008:HFS_Pstates_mup}
\begin{equation}
 \delta = 0.14456 ~ \textrm{meV}
\label{eq:delta}
\end{equation}
as shown in Fig.~\ref{fig:energy-levels}.
The values given in Table~8 of~\cite{Borie:2012:LS_revisited} include
this shift and deviate by less than 0.0002\,meV from the values in
Eqs.~(\ref{eq:hfs_2P}) and (\ref{eq:delta}).

$\Delta E_\mathrm{L}$ and $\Delta E_\mathrm{HFS}$ can be deduced
independently of each other from
\begin{equation}
\label{eq:linear-combination}
\begin{split}
 \frac{1}{4} h\nu_s + \frac{3}{4} h\nu_t
        =& \Delta E_\mathrm{L} + \Delta E_\mathrm{FS} 
                + \frac{1}{8}\Delta E_\mathrm{HFS}^{2\mathrm{P}_{3/2}} 
                + \frac{1}{4}\delta \\
        =& \Delta E_\mathrm{L} + 8.8123(2) ~ \textrm{meV} \\[5mm]
 h\nu_s - h\nu_t 
        =& \Delta E_\mathrm{HFS}
                - \Delta E_\mathrm{HFS}^{2\mathrm{P}_{3/2}} + \delta \\
        =& \Delta E_\mathrm{HFS} - 3.2480(2) ~ \textrm{meV}
\end{split}
\end{equation}
where the uncertainties of the constant terms correspond to
uncalculated higher-order QED terms, and differences between the
various authors.  These uncertainties are small compared to
both the uncertainties of the measured transition
frequencies~\cite{Antognini:2012:Nature_mup2} and the uncertainties of
the theoretical prediction for $ \Delta E_\mathrm{L}$ and $\Delta
E_\mathrm{HFS}$ (see below).

Finite-size effects are small for the 2P states but have been
included in the theoretical prediction of both the fine and hyperfine
contributions.  For the fine splitting they have been computed
perturbatively to be 
$-0.0000519\,\rpC^2=-0.00004$~meV~\cite{Borie:2012:LS_revisited}
and with an all-order approach
$-0.0000521\,\rpC^2=-0.00004$~meV~\cite{Indelicato:2012}.  The
finite-size contributions to the $2\mathrm{P}_{3/2}$ hyperfine splitting are
$<10^{-5}$~meV~\cite{Indelicato:2012}.
Hence the finite-size related uncertainties of the constant terms in
Eqs.~(\ref{eq:linear-combination}) are 
negligible at the present level of experimental accuracy.

\section{The Lamb shift prediction and the proton charge radius}
\label{sec:lamb-shift_rp}

The main contribution to the Lamb shift in muonic hydrogen
is given by the one-loop electron-positron vacuum
polarization (eVP).
The second largest term is due to the finite extension of the
proton charge and is proportional to $m_r^3\, \rpC^2$, where $m_r
\approx 186m_e$ is the reduced mass of the $\mu$p system 
and $m_e$ the electron mass.
The third largest contribution is given by the
K\"all$\acute{\mathrm{e}}$n-Sabry (two-loop eVP) diagrams.
The fourth largest term is
the muon one-loop self-energy summed with the one-loop
muon-antimuon vacuum polarization ($\mu$VP).

In has to be stressed that the observed discrepancy (Eq.~\ref{eq:discrepancy}) 
is larger than any contribution except for the four terms listed above.
Therefore a solution of the proton radius puzzle in the context of
muonic hydrogen theory could arise either from one-loop VP, or proton
structure effects, or fundamental problems in bound-state QED, but hardly
from a missing or wrong higher-order contribution.
Nevertheless it is necessary to compute all these higher order
contributions to better than $\sim 0.001$~meV in order to exploit the
accuracy of the measurements and to determine \rpC{} with a relative
accuracy of $u_r\approx 3\times 10^{-4}$, which corresponds to 0.0003~fm.
The radiative, relativistic, binding and recoil corrections to $\Delta
E_\mathrm{L}$ are summarized in Table~\ref{tab:muonic-h-lamb},
whereas the proton-structure dependent contributions to $\Delta
E_\mathrm{L}$ are given in
Table~\ref{tab:fs}.

\subsection{Proton structure independent contributions}

The corrections to the energy levels predicted by the Schr\"odinger
equation solved for the point-charge Coulomb potential are usually 
calculated using
perturbation theory.
Relativistic corrections are obtained from the two-body
Breit-Hamiltonian.
To check the validity of the perturbative approach the largest
contribution (one-loop eVP) has been recalculated by 
numerical integration of the Dirac
equation~\cite{Indelicato:2012,Borie:2012:LS_revisited,
Carroll:2011:FF_dependence,Carroll:2011:NP}.
The inclusion of the Uehling term to the Coulomb potential yields the
relativistic all-order one-loop eVP correction given by the sum of items
\#3 (205.02821 meV) and \#5 (0.15102 meV) in Table~\ref{tab:muonic-h-lamb}.
This compares well with the perturbative results: \#1 (205.0074 meV) and the corrections for 
relativistic and eVP iterations given by \#2 (0.018759~meV) and \#5 (0.1507~meV).
Item \#5 is the energy shift in second order perturbation theory
related to the wave function distortion caused by one-loop eVP.
Item \#7 is a similar correction to the K\"all$\acute{\mathrm{e}}$n-Sabry
contribution. Both originate from the wave function distortion
caused by one-loop eVP.

The relativistic eVP correction with the full reduced mass
dependence was given in~\cite{Jentschura:2011:relrecoil}. 
It amounts to $0.018759$~meV.
Karshenboim {\it et al.} found~\cite{Karshenboim:2012:relrecoil}.
why this differs from the previous result of 0.0169~meV given in item \#2:
The previous work had been done in different gauges, and
in some of those gauges some contributions (retardation and
two-photon-exchange effects) had been forgotten.
In \#19 we give the difference between the full expansion in ($m/M$)
of the relativistic eVP correction (Eq.~(4)
in~\cite{Karshenboim:2012:relrecoil}) and its lowest order (Eq.~(6)
in~\cite{Karshenboim:2012:relrecoil}):
$E_\mathrm{VP}^\mathrm{(rel)}(2\mathrm{P}_{1/2}-2\mathrm{S}_{1/2})-E_\mathrm{VP}^\mathrm{(0)}(2\mathrm{P}_{1/2}-2\mathrm{S}_{1/2})=0.018759-0.020843=-0.002084$~meV.
If one includes these corrections, the various approaches for the one-loop
eVP contributions with relativistic-recoil corrections are consistent:
Pachucki, who starts from the Schr\"odinger equation, becomes
$205.0074+0.018759=205.0262$~meV.
Borie, who starts from the Dirac
equation, becomes $205.0282-0.002084=205.0261$~meV.

Item \#11, the muon self-energy correction to eVP of order
$\alpha^2(Z\alpha)^4$, was improved as summarized in Eq.~(27)
of~\cite{Jentschura:2011:SemiAnalytic}. It includes
contributions
beyond the logarithmic term with modification of the Bethe logarithm
to the Uehling potential.

Item \#12, eVP loop in self-energy, is part of \#21 as can be seen
from Fig.~22 in~\cite{BorieRinker:1982:muAtoms}. Thus it had erroneously been
double-counted in Ref.~\cite{Pohl:2010:Nature_mup1_suppl} (i.e.\ Ref.~\cite{Pohl:2010:Nature_mup1}).


\begin{table*}[h!]
\renewcommand{\baselinestretch}{1.1}
\renewcommand{\arraystretch}{1.5}
  \caption{All known radius-{\bf independent} contributions to the
    Lamb shift in $\mu$p from different authors, and the one we
    selected (usually the all-order
    calculations which we consider more complete).
    Values are
    in meV.  The entry \# in the first column refers to Table 1 in
    Ref.~\cite{Pohl:2010:Nature_mup1_suppl}.  The "finite-size to
    relativistic recoil correction" (entry \#18 in
    ~\cite{Pohl:2010:Nature_mup1_suppl}) which depends on the proton
    structure has been shifted to Table~\ref{tab:fs}, together with
    the small terms \#26 and \#27, and the proton polarizability term
    \#25.\\ SE: self-energy, VP: vacuum polarization, LBL:
    light-by-light scattering, Rel: relativistic, NR: non-relativistic,
    RC: recoil correction.}
  \label{tab:muonic-h-lamb}
\setlength\tabcolsep{1mm}
\setlength{\extrarowheight}{0.5mm}
{
  \fontsize{6pt}{6pt}\selectfont
  \begin{tabular}{r | l |f{7} f{8} f{7} f{9} f{9} l}
  \hline
  \hline
\# & Contribution                                            & \cnt{1}{Pachucki}
                                                                             & \cnt{1}{Nature}
                                                                                            & \cnt{1}{Borie-v6}
                                                                                                          & \cnt{1}{Indelicato}
                                                                                                                            & \cnt{1}{Our choice}
                                                                                                                                             & Ref. \\
   &                                                         & \cnt{1}{~\cite{Pachucki:1996:LSmup,Pachucki:1999:ProtonMup}}
                                                                            & \cnt{1}{\cite{Pohl:2010:Nature_mup1_suppl}} 
                                                                                            & \cnt{1}{\cite{Borie:2012:LS_revisited}}
                                                                                                          & \cnt{1}{\cite{Indelicato:2012}}\\
\hline
 1 & NR one-loop electron VP (eVP)                           &  205.0074    &               &             &                 &                 &      \\
 2 & Rel. corr. (Breit-Pauli)                                &    0.0169\footnote{This value has been recalculated to be 0.018759~meV~\cite{Jentschura:2011:relrecoil}.}  &               &             &                &  \\
 3 & Rel. one-loop eVP                                       &              &  205.0282     &  205.0282   &  205.02821      &  205.02821      &  \cite{Indelicato:2012} Eq.(54) \\
19 & Rel. RC to eVP,   $\alpha(Z\alpha)^4$                   & \ttxt{(incl. in \#2)\footnote{This correction is not necessary here because in \#2 the Breit-Pauli contribution has been calculated using a Coulomb potential modified by eVP. }}
                                                                            &   -0.0041     &   -0.0041   &                 & -0.00208\footnote{Difference between Eq. (6) and (4) in~\cite{Karshenboim:2012:relrecoil}: $E_\mathrm{VP}^\mathrm{(rel)}(2\mathrm{P}_{1/2}-2\mathrm{S}_{1/2})-E_\mathrm{VP}^\mathrm{(0)}(2\mathrm{P}_{1/2}-2\mathrm{S}_{1/2})=0.018759-0.020843=-0.002084$~meV (see also Table IV). 
 Using these corrected values, the various approaches are consistent.   Pachucki becomes $205.0074+0.018759=205.0262$~meV and  Borie  $205.0282-0.0020843=205.0261$~meV.}
                                                                                                                                              &  \cite{Jentschura:2011:relrecoil, Karshenboim:2012:relrecoil} \\
\hline
 4 & Two-loop eVP (K\"all$\acute{\mathrm{e}}$n-Sabry)        &    1.5079    &    1.5081     &    1.5081   &    1.50810      &    1.50810      &  \cite{Indelicato:2012} Eq.(57) \\
\hline 
 5 & One-loop eVP in 2-Coulomb                               &    0.1509    &    0.1509     &    0.1507   &    0.15102      &    0.15102      &  \cite{Indelicato:2012} Eq.(60) \\
   & \quad  lines $\alpha^2(Z\alpha)^5$                      \\
 7 & eVP  corr.  to   K\"all$\acute{\mathrm{e}}$n-Sabry      &    0.0023    &    0.00223    &    0.00223  &    0.00215      &    0.00215      &  \cite{Indelicato:2012} Eq.(62), \cite{Ivanov:2009:three-loop}  \\[2mm]
 6 & NR three-loop eVP                                       &    0.0053    &    0.00529    &    0.00529  &                 &    0.00529      &  \cite{Kinoshita:1999-three-loop, Ivanov:2009:three-loop}          \\
\hline
 9 & Wichmann-Kroll, ``1:3'' LBL                             &              &   -0.00103    &   -0.00102  &   -0.00102      &   -0.00102      &  \cite{Indelicato:2012} Eq.(64), \cite{Karshenboim:2010:JETP_LBL}   \\
10 & Virtual Delbr\"{u}ck, ``2:2'' LBL                       &              &    0.00135    &    0.00115  &                 &    0.00115      &  \cite{Eides:2001:HydTheo, Karshenboim:2010:JETP_LBL}  \\
\!\!new  & ``3:1'' LBL                                       &              &               &   -0.00102  &                 &   -0.00102      &  \cite{Karshenboim:2010:JETP_LBL}   \\
\hline

20 & $\mu$SE and $\mu$VP                                     &  -0.6677     &   -0.66770    &   -0.66788  &   -0.66761      &   -0.66761      &  \cite{Indelicato:2012} Eqs.(72)+(76)\\
\hline
11 & Muon SE corr. to eVP  $\alpha^2(Z\alpha)^4$              &    -0.005(1) &   -0.00500    &   -0.004924~\footnote{In Appendix C, incomplete.} 
                                                                                                          &                 &   -0.00254      &  \cite{Jentschura:2011:SemiAnalytic} Eq.(29a)~\footnote{Eq. (27) in~\cite{Jentschura:2011:SemiAnalytic} includes contributions beyond the logarithmic term with  modification of the Bethe logarithm to the Uehling potential.  The factor 10/9 should be replaced  by 5/6.  } \\[2mm]
12 & eVP loop in self-energy  $\alpha^2(Z\alpha)^4$           &    -0.001    &   -0.00150    &             &                &   \footnote{This term is part of \#22, see Fig.~22 in~\cite{BorieRinker:1982:muAtoms}.}
                                                                                                                                              &  \cite{Eides:2001:HydTheo, Barbieri:1973,Suura:1957,Petermann:1957}  \\
21 & Higher-order corr. to  $\mu$SE and $\mu$VP              &              &   -0.00169    &   -0.00171~\footnote{Borie includes wave function corrections calculated in \cite{Ivanov:2009:three-loop}. The actual difference between Ref.~\cite{Pohl:2010:Nature_mup1_suppl} and Borie-v6~\cite{Borie:2012:LS_revisited} is given by the inclusion of the K\"all$\acute{\mathrm{e}}$n-Sabry correction with muon loop.}  
                                                                                                          &                 &   -0.00171      &  \cite{BorieRinker:1982:muAtoms} Eq.(177)  \\[2mm]
13 & Mixed eVP+$\mu$VP                                       &              &    0.00007    &    0.00007  &                 &    0.00007      &  \cite{Eides:2001:HydTheo}                              \\
\!\!new & eVP and $\mu$VP in two Coulomb lines               &              &               &             &    0.00005      &    0.00005      &  \cite{Indelicato:2012} Eq.(78) \\
\hline
14 & Hadronic VP $\alpha(Z\alpha)^4m_r$                      &    0.0113(3) &    0.01077(38)&    0.011(1) &                  &    0.01121(44)  &  \cite{Friar:1999,Martynenko:2000,Martynenko:2001}              \\
15 & Hadronic VP $\alpha(Z\alpha)^5m_r$                      &              &    0.000047   &             &                  &    0.000047     &  \cite{Martynenko:2000,Martynenko:2001} \\
16 & Rad corr. to hadronic VP                                &              &   -0.000015   &             &                 &   -0.000015    &  \cite{Martynenko:2000,Martynenko:2001}  \\
\hline
17 & Recoil corr.                                            &   0.0575     &    0.05750    &    0.0575   &    0.05747      &    0.05747     &   \cite{Indelicato:2012} Eq.(88) \\
22 & Rel. RC $(Z\alpha)^5$                                   &  -0.045      &   -0.04497    &   -0.04497  &   -0.04497      &    -0.04497     &  \cite{Indelicato:2012} Eq.(88), \cite{Eides:2001:HydTheo}\\
23 & Rel. RC $(Z\alpha)^6$                                   &   0.0003     &    0.00030    &             &    0.0002475    &     0.0002475   &  \cite{Indelicato:2012} Eq.(86)+Tab.II \\
\!\!new & Rad. (only eVP) RC $\alpha(Z\alpha)^5$             &              &               &             &                 &     0.000136     &  \cite{Jentschura:2011:SemiAnalytic} Eq.(64a) \\
\hline
24 & Rad. RC $\alpha(Z\alpha)^n$ (proton SE)                 &  -0.0099     &   -0.00960    &   -0.0100   &                 &    -0.01080(100) &  \cite{HillPaz:2011:TwoPhotons}\footnote{This was calculated in the framework of NRQED. It is related to the definition of the proton radius.} \cite{Eides:2001:HydTheo}                \\
\hline
   & Sum                                                     & 206.0312     &  206.02915    &  206.02862  &                 &   206.03339(109) &            \\
\hline
  \hline
\end{tabular}
}
%
\end{table*}
%
%

\subsection{Proton structure dependent contributions}

We first present the background required to understand the individual
proton structure dependent contributions summarized in
Table~\ref{tab:fs}.
The main contribution due to finite nuclear size has been given
analytically to order $(Z\alpha)^6$ by
Friar~\cite{Friar:1978:Annals}. The main result
is~\cite{Borie:2012:LS_revisited}
\begin{equation}
\label{eq:friar}
\Delta E_\mathrm{finite\;size}=-\frac{2 \pi \alpha }{3} |\Psi(0)|^2 \left[\;
  \langle r^2 \rangle -\frac{ Z \alpha  }{2}\,m_r\,\rZt
		     +(Z \alpha )^2 \left(F_\mathrm{REL}+m_r^2 F_\mathrm{NREL}\right)\right]
\end{equation}
with
\begin{equation}
\label{eq:friar-2}
\begin{array}{ll}
F_\mathrm{REL} & = -\langle r^2 \rangle \left[ \gamma-\dfrac{35}{16}
               + \ln(Z\alpha)+\langle \ln(m_r r)\rangle  \right]
               -\dfrac{1}{3}\langle r^3 \rangle\langle \dfrac{1}{r}\rangle
               +  I_2^\mathrm{REL} + I_3^\mathrm{REL}\\[5mm]
F_\mathrm{NREL}& =\dfrac{2}{3} (\langle r^2 \rangle)^2 \left[ \gamma-\dfrac{5}{6}+\ln(Z\alpha)\right]
             +\dfrac{2}{3}\langle r^2 \rangle \langle r^2 \ln(m_r r)\rangle  
             -\dfrac{\langle r^4 \rangle}{40}+ \langle r^3 \rangle \langle r \rangle  \\[2mm]
 
             &+ \dfrac{1}{9}\langle r^5 \rangle \langle \dfrac{1}{r} \rangle
             + I_2^\mathrm{NREL} + I_3^\mathrm{NREL}
\end{array}
\end{equation}
where $\langle r^n \rangle=\int \mathrm{d}^3r \, r^n \rho(\mbox{\boldmath{$r$}})$ is the $n^\mathrm{th}$ moment
of the charge distribution, $\rpC{}=\sqrt{\langle r^2 \rangle}\approx0.84$~fm
is the rms charge radius, \rZt{} the third Zemach moment (see below),
$m_r$ the reduced mass, $\gamma$ the Euler's constant, and
$I_2^\mathrm{NREL}$, $I_3^\mathrm{NREL}$, $I_2^\mathrm{REL}$ and
$I_3^\mathrm{REL}$ are integrals which depend on the charge distributions~\cite{Friar:1978:Annals}.
%
%

\subsubsection{One-photon exchange contribution}
The coefficient $b_a=-\frac{2 \pi \alpha }{3} |\Psi(0)|^2$ of the
first term in Eq.~(\ref{eq:friar}) describes the leading
finite-size effect (for the sake of comparison we follow the notation
$b_x$ given in Table B.14 of~\cite{Borie:2012:LS_revisited}).
For the 2S state
$b_a=-5.1973$~meV/fm$^2$~\cite{Borie:2012:LS_revisited} calculated
using the Schr\"odinger wave function $\Psi$. 

Relativistic corrections
to the finite-size effect are accounted for in $F_\mathrm{REL}$. 
The coefficient proportional to $\langle r^2 \rangle$ in
$F_\mathrm{REL}$ is 
\begin{equation}
b_c=\frac{2 \pi \alpha }{3} |\Psi(0)|^2 \,(Z \alpha)^2 \left[
  \gamma-\dfrac{35}{16} + \ln(Z\alpha)
  \right].
\end{equation}
For the 2S-state $b_c=-0.00181$~meV/fm$^2$.
The total one-photon exchange contribution in this then given by
\begin{equation}
\label{eq:ope_main_1}
\Delta E_\mathrm{OPE} =
 b_a + b_c =
 -5.1973\rr{} - 0.00181\rr{}  ~ \mathrm{meV} =
 -5.1991\rr{} ~ \mathrm{meV}.
\end{equation}
The complete, relativistic, all-order calculations~\cite{Indelicato:2012} obtain
\begin{equation}
\label{eq:ope_main}
\Delta E_\mathrm{OPE} =  -5.1994\rr{} ~ \mathrm{meV}
\end{equation}
which we will use (see Tab.~\ref{tab:fs}).
For the 2P$_{1/2}$ state the corresponding finite size terms amount to 
$-0.0000519$~meV/fm$^2$~\cite{Borie:2012:LS_revisited}.

\subsubsection{Third Zemach moment contribution}

The second term in Eq.~(\ref{eq:friar}) amounts to 
\begin{equation}
\label{eq:91}
\Delta E_\mathrm{third\; Zemach}=0.0091 \rZt  ~ \mathrm{meV}. 
\end{equation}
This is the second largest finite-size effect and it depends on the 
third Zemach moment
\begin{equation}
\label{eq:thirdZemach}
 \rZt{} = \int \mathrm{d}^3r \int \mathrm{d}^3r' \; {r}^3 \,\rho_E
(\mbox{\boldmath{$r$}}-\mbox{\boldmath{$r'$}})\rho_E (\mbox{\boldmath{$r'$}})
\end{equation}
which can be determined experimentally from the electric Sachs form
factor $G_E$ measured in elastic electron-proton scattering
as~\cite{FriarSick:2005}
\begin{equation}
\label{eq:third-zemach-form-factor}
\rZt{}= \frac{48}{\pi}\int\dfrac{dQ}{Q^4}[G^2_E(Q^2)-1+\dfrac{1}{3}Q^2\langle r ^2 \rangle ]\,.
\end{equation}
Commonly used values are
$\rZt{}=2.71(13)$~fm$^3$~\cite{FriarSick:2005} and
$\rZt{}=2.85(8)$~fm$^3$~\cite{Distler:2011:Zemach}.
It is customary~\cite{Pachucki:1999:ProtonMup, Borie:2012:LS_revisited, Indelicato:2012} to express the third Zemach moment using the second
moment of the charge distribution as
\begin{equation}
\label{eq:f-factor}
\rZt{}=f \langle r^2 \rangle^{3/2}
\end{equation}
where $f$ is a constant which depends on the model for the shape of
the proton.
For an exponential charge distribution (dipole form factor) $f=3.79$.
For a Gaussian distribution $f=3.47$.
Values extracted from the measured electric form factors are
$f=3.78(31)$~\cite{FriarSick:2005} and
$f=4.18(13)$~\cite{Distler:2011:Zemach}.

Adopting $f=4.0(2)$ from Ref.~\cite{Borie:2012:LS_revisited} to account for the
spread of the various values measured in scattering experiments, one gets
from Eq.~(\ref{eq:91}) and  Eq.~(\ref{eq:f-factor})
\begin{equation}
\label{eq:r^3}
\Delta E_\mathrm{third\; Zemach}=0.0365(18)\langle r^2 \rangle^{3/2} ~ \mathrm{meV}
\end{equation}
(see first column in Table~\ref{tab:fs}).

A solution of the proton radius puzzle assuming a large tail of the
proton charge distribution resulting in an extremely large \rZt{}
value~\cite{DeRujula:2010:NotEndangered, DeRujula:2011:QEDconfronts},
and hence a large value of $f$, 
has been ruled out by electron-proton scattering
data~\cite{FriarSick:2005,
  CloetMiller:2011:3rdZemach,Distler:2011:Zemach} and by chiral
perturbation theory~\cite{Pineda:2005:Lamb-shift:proton-radius-def,
  Pineda:2012}.

The third Zemach contribution may be seen simplistically as a
second-order correction in perturbation theory given by the
modification of the wave function caused by the finite size.
In a quantum field framework it is part of the two-photon exchange
diagrams~\cite{CarlsonVanderhaeghen:2011, Jentschura:2011:AnnPhys1,
  Pachucki:1996:LSmup, Pineda:2005:Lamb-shift:proton-radius-def}.

\subsubsection{Two-photon exchange contributions}

The two-photon contributions with finite-size are usually divided
into an elastic part,
where the intermediate virtual proton remains on-shell, and an
inelastic part (proton polarizability contribution $\Delta
E_\mathrm{pol}$), where the virtual proton is off-shell.
The elastic part is approximately given by the \rZt{} term in Eq.~(\ref{eq:friar}).

A unified treatment of such contributions can be achieved in modern
quantum field theory using the doubly-virtual Compton amplitude which
can be related to measured form factors and spin averaged structure
functions using dispersion relations.
Part of a subtraction term needed to remove a divergence in one
Compton amplitude is usually approximated using the one-photon
on-shell form factor~\cite{Pachucki:1999:ProtonMup}.
A possible large uncertainty related to this approximation has been
emphasized by Miller {\it et al.}~\cite{Miller:2011:NaturalResolution}
as well as Hill and Paz~\cite{HillPaz:2011:TwoPhotons}.
However, Birse and McGovern calculated this subtraction term in the
framework of heavy baryon chiral perturbation
theory~\cite{BirseMcGovern:2012}. 
They obtain a contribution of $\Delta E_\mathrm{sub}=-0.0042(10)$\;meV,
in agreement with~\cite{Pachucki:1996:LSmup,Pachucki:1999:ProtonMup,
Martynenko:2006:Pol_mup_H,CarlsonVanderhaeghen:2011}.
This value is about two orders of magnitude smaller than the discrepancy.

The total two-photon exchange contribution with finite-size amounts to~\cite{Pachucki:1999:ProtonMup,
  CarlsonVanderhaeghen:2011,BirseMcGovern:2012,
  Jentschura:2011:AnnPhys1}
\begin{equation}
\label{eq:tpe}
\Delta E_\mathrm{TPE} = 0.0332(20) ~ \mathrm{meV}.
\end{equation}
It results from the sum of an elastic part $\Delta
E_\mathrm{el}=0.0295(16)$~meV~\cite{CarlsonVanderhaeghen:2011, BirseMcGovern:2012}, a
non-pole term $\Delta
E_\mathrm{np}=-0.0048$~meV~\cite{BirseMcGovern:2012,CarlsonVanderhaeghen:2011},
the subtraction term $\Delta
E_\mathrm{sub}=-0.0042(10)$~meV~\cite{BirseMcGovern:2012} and an
inelastic (proton polarizability) contribution $ \Delta
E_\mathrm{pol}=0.0127(5)$~meV~\cite{CarlsonVanderhaeghen:2011}.
Additional contributions from ``new models'' for the ``off-shell''
form factor~\cite{Miller:2012:quasi-elastic} have been limited to
$<0.001$~meV by quasi-elastic electron
scattering~\cite{BirseMcGovern:2012}.
The uncertainty of $\Delta
E_\mathrm{el}$
accounts for various measured form factors (Kelly~\cite{Kelly:2004},
AMT~\cite{Arrington:2007} and Mainz
2010~\cite{Bernauer:2010:NewMainz, Vanderhaeghen:2010}).
The sum
\begin{equation}
\label{eq:sum-el-np-sub}
\Delta E_\mathrm{el}+\Delta E_\mathrm{np}=0.0247(16) ~ \mathrm{meV}
\end{equation}
should be compared to $\Delta E_\mathrm{third\; Zemach}$ corrected
for recoil corrections~\cite{Jentschura:2011:AnnPhys1} which reduce
the third Zemach contribution by a factor
(0.018~meV)/(0.021~meV)~\cite{Pachucki:1996:LSmup} to
\begin{equation}
\label{eq:zemach-recoil}
\Delta E_\mathrm{third\; Zemach+recoil}\approx \frac{0.018}{0.021}\cdot 0.0091 \rZt{} =0.0217(15) ~ \mathrm{meV} .
\end{equation}
The uncertainty arises from the recoil corrections ($18/21=0.86(4)$)
and the use of  $\rZt{}= 2.78(14)$~fm$^3$ obtained 
from the spreads of the values in~\cite{FriarSick:2005, Distler:2011:Zemach}.
Comparison between Eq.~(\ref{eq:sum-el-np-sub}) and
Eq.~(\ref{eq:zemach-recoil}) shows that there is fair agreement
between the results from 
Eq.~(\ref{eq:friar}) and the more advanced quantum field theoretical approach 
via Compton scattering and dispersion relations described above.
The approximate scaling of $0.018/0.021$ has been discussed by
Jentschura~\cite{Jentschura:2011:AnnPhys1} based on
Pachucki~\cite{Pachucki:1996:LSmup}, to account for recoil corrections
which are automatically included in the unified treatment of the
two-photon exchange contribution $\Delta E_\mathrm{TPE}$.
This recoil correction to two-photon finite-size contributions is
$\sim -0.003$~meV, much less and of opposite sign compared to the term
of 0.013~meV (\#18 in Table~\ref{tab:fs}) given by Borie who follows
Friar.

\subsubsection{Higher-order moments of the charge distribution}

The sum of all terms in Eq.~(\ref{eq:friar}) beyond $\langle r^2
\rangle$ and \rZt{}, which have been calculated assuming an
exponential charge distribution and \rpC{}=0.875~fm, affects the Lamb
shift by $\Delta E_\mathrm{finite\; size}^\mathrm{ho, \;
  Borie}=-0.000123$~meV~\cite{Borie:2012:LS_revisited}.
Even though the assumption of an exponential distribution may not be
completely realistic for these higher-order contributions, this is
sufficient~\cite{Borie:2012:LS_revisited}.
The smallness of this term may be qualitatively understood in a
perturbative framework (Eq.~(\ref{eq:friar})): higher
moments of the charge distribution are always scaled by higher
powers of $(Z\alpha)$.

\begin{table*}[t!]
\renewcommand{\baselinestretch}{1.1}
\renewcommand{\arraystretch}{1.5}
\setlength\tabcolsep{1mm}
\footnotesize
\caption{Proton structure dependent contributions to the Lamb shift in
  $\mu$p from different authors, and the one we selected (usually the all-order
    calculations which we consider more complete).  Values are
  in meV, \rr{} in fm$^2$.  The
  entry \# in the first column refers to Table 1 in Ref.~\cite{Pohl:2010:Nature_mup1_suppl}
  supplementary informations~\cite{Pohl:2010:Nature_mup1}.  Entry
  \#~18 is under debate. \\TPE: two-photon exchange, VP:  vacuum polarization, SE: self-energy, Rel: relativistic. }
\label{tab:fs}
{
  \fontsize{6pt}{6pt}\selectfont
  \begin{tabular}{r|l|lll ll l }
\hline
\hline
\# & Contribution                                           & Borie-v6         & Karshenboim   & Pachucki  & Indelicato & Carroll      & Our choice    \\
   &                                                        & \cite{Borie:2012:LS_revisited}
                                                                               & \cite{Karshenboim:2012:relrecoil} & \cite{Pachucki:1996:LSmup,Pachucki:1999:ProtonMup} & \cite{Indelicato:2012} &\cite{Carroll:2011:FF_dependence}\\

\hline
 &  Non-rel.  finite-size                                   &  -5.1973\rr{}       &  -5.1975\rr{}  & -5.1975\rr{}                                                       
                                                                                                                     &              &              &           \\
 &  Rel. corr. to non-rel. finite size                      &  -0.0018\rr{}       &                &  -0.0009 meV\footnote{Corresponds to Eq. (6) in \cite{Pachucki:1999:ProtonMup} which accounts only for the main terms in $F_\mathrm{REL}$ and $F_\mathrm{NREL}$.}    &              &              &           \\[1mm]
 &  Rel. finite-size                                        &                     &                &                  &              &              &           \\
 &  $\quad $ exponential                                    &                     &                &                  & -5.1994\rr{} & -5.2001\rr{} &  -5.1994\rr{} \\
 &  $\quad $ Yukawa                                         &                     &                &                  &              & -5.2000\rr{} &           \\
 &  $\quad $ Gaussian                                       &                     &                &                  &              & -5.2001\rr{} &           \\
   
\hline 
 &  Finite size corr. to one-loop eVP                       & -0.0110\rr{}        & -0.0110\rr{}   & -0.010\rr{}     & -0.0282\rr{} &              & -0.0282\rr{}     \\
 &  Finite size  to one-loop eVP-it.                        & -0.0165\rr{}        & -0.0170\rr{}   & -0.017\rr{}      & \!\!(incl. in -0.0282)      &&                  \\
 &  Finite-size corr. to K\"all$\acute{\mathrm{e}}$n-Sabry                       & \footnote{This contribution has been accounted already in both the -0.0110~meV/fm$^2$ and -0.0165~meV/fm$^2$  coefficients.}  &                &                  & -0.0002\rr{} &              & -0.0002\rr{}     \\
\!\!new &  Finite size corr. to $\mu$ self-energy                   & (0.00699)~\footnote{Given only in  Appendix C. Bethe logarithm is not included.}&&                  & 0.0008\rr{} &              & 0.0009(3)\rr{}\footnote{This uncertainty accounts for the difference between all-order in $Z\alpha$ and perturbative approaches~\cite{Indelicato:Mohr:2012}.}   \\   
\hline
 &  $\Delta E_\mathrm{TPE}$~\cite{BirseMcGovern:2012} & &                &                  &                &              & 0.0332(20) meV   \\[1mm]
 &  elastic (third Zemach)\footnote{Corresponds to Eq.~(\ref{eq:sum-el-np-sub}). }
                                                            &                     &                &                  &                &              &                 \\
 &  $\quad $ measured \rZt                                  &  0.0365(18)\rrr{}   &                &                  &                &              & (incl. above)   \\
 &  $\quad $ exponential                                    &                     &                &\!\!0.0363\rrr{}  &   0.0353\rrr{}\;\footnote{This value is slightly different from Eq.~(\ref{eq:fs-paul}) because here an all-order in finite-size AND an all-order in eVP approach was used.}
                                                                                                                                       & 0.0353\rrr{} &                 \\
 &  $\quad $ Yukawa                                         &                     &                &                  &                & 0.0378\rrr{} &                 \\
 &  $\quad $ Gaussian                                       &                     &                &                  &                & 0.0323\rrr{} &                 \\
 
25 &  inelastic (polarizability)              & 0.0129(5) meV~\cite{CarlsonVanderhaeghen:2011} &  &\!\!0.012(2) meV      &      &              &   (incl. above)    \\
 
 \hline
\!\!new &  Rad. corr. to TPE                            & -0.00062\rr{} &       &                &                 &                &  -0.00062\rr{}         \\[2mm]

26  &  eVP corr. to polarizability                           &                &                &                 &             &                &0.00019 meV~\cite{Martynenko:2001}       \\
27 &  SE corr. to polarizability                         &                 &                &                 &              &                &-0.00001 meV~\cite{Martynenko:2001}        \\
                              
\hline
18 &  Finite-size to rel. recoil corr.                        & (0.013 meV)~\footnote{See Appendix F of~\cite{Friar:1978:Annals}. This term is under debate.}           &                 & \footnote{Included in $\Delta E_\mathrm{TPE}$. This correction of $0.018-0.021=-0.003$ meV is given by  Eq.~(64) in \cite{Pachucki:1996:LSmup} and  Eq.~(25)~in \cite{Pachucki:1999:ProtonMup}. This correction is also discussed in~\cite{Jentschura:2011:AnnPhys1} where the 6/7 factor results from 0.018/0.021.  }  &     &                & (incl. in $\Delta E_\mathrm{TPE}$) \\  
\hline
 &  Higher-order finite-size corr.      & -0.000123~meV &                 &                 &\!\!0.00001(10)~meV           &                &0.00001(10)~meV \\
\hline
 &  $2\mathrm{P}_{1/2}$ finite-size corr.          &  -0.0000519\rr{}~\footnote{Eq.~(6a) in~\cite{Borie:2012:LS_revisited}. }     &                 &                 & (incl. above)& (incl. above) &     (incl. above)\\
\hline
\hline
\end{tabular}
}
\end{table*}

The dependence of the $\langle r^2 \rangle$ coefficient on the assumed
proton charge distribution has been shown to be
weak~\cite{Borie:2005:LSmup, Carroll:2011:FF_dependence}.  This was
demonstrated by numerical integration of the Dirac equation using
various proton charge distributions: exponential, Gaussian and Yukawa
(see Table~\ref{tab:fs}).
Indelicato~\cite{Indelicato:2012} determined the total finite-size effect,
also by numerical integration of the Dirac equation,
using a dipole charge distribution (Eq.~(44) in \cite{Indelicato:2012})
\begin{equation}
\label{eq:fs-paul}
\begin{array}{ll}
\Delta E_\mathrm{finite\; size}=& -5.19937\rpC{}^2  + 0.03466\rpC{}^3 + 0.00007\rpC{}^4\\
                              & -0.000017\rpC{}^5 + 1.2\cdot10^{-6}\rpC{}^6 + 0.00027\rpC^2\log{(\rpC)} ~ \textrm{meV} .
\end{array}
\end{equation}
Equation~(\ref{eq:fs-paul}) was attained by fitting the eigenvalues of
the Dirac equation obtained for a finite-size Coulomb potential for
various values of the proton charge radius.
Hence, it accounts for all-order finite-size effects.
%

The first coefficient of this equation is in agreement with
$b_a+b_c=-5.1973-0.00181=-5.1991$~meV/fm$^2$ of
\cite{Borie:2012:LS_revisited} and the second one is compatible with
Eq.~(\ref{eq:r^3}).
This implies that the two approaches, one starting from the Dirac
equation with finite-size-corrected Coulomb potential, and the other one
starting from the Schr\"odinger solution (with point-like Coulomb
potential) complemented with relativistic and finite-size corrections,
are equivalent.
The sum of the terms of Eq.~(\ref{eq:fs-paul}) beyond $\rpC^2$ and
$\rpC^3$ is only $\sim0.00004$~meV, suggesting that the higher moments
of the charge distribution do not affect significantly the prediction
of the muonic hydrogen Lamb shift.

Sick~\cite{FriarSick:2005} showed that \rZt{} extracted from the integral of
Eq.~(\ref{eq:third-zemach-form-factor}) applied at the world data is most sensitive to the cross
sections measured at $Q^2\approx 0.05$~(GeV/c)$^2$.
Higher $Q^2$ data contributes, too~\cite{DeRujula:2011:QEDconfronts}, 
but \rZt{} is rather insensitive to the lowest $Q^2$ region.
Therefore \rZt{} cannot be dramatically increased by 
contributions from very low $Q^2$, where no data is 
available~\cite{CloetMiller:2011:3rdZemach}.

The theoretical prediction of the finite-size terms given in
Eq.~(\ref{eq:friar}) could potentially become problematic if the higher moments
of the charge distribution $\langle r^4 \rangle$, $\langle r^5
\rangle$, $\langle r^6
\rangle$ and $\langle r^2 \rangle\langle \log{r} \rangle$ etc. were
large.
However, these moments have been evaluated using electron-scattering
data~\cite{Distler:2011:Zemach} down to the lowest experimentally
accessible exchanged photon momentum of
$Q^2_\mathrm{min}=0.004$~(GeV/c)$^2$.
The values reported in~\cite{Distler:2011:Zemach} are small
 suggesting that the finite-size effect in muonic
hydrogen is properly described by the $\langle r^2 \rangle$ and \rZt{}
terms alone.

Extending De R\'ujula's argument~\cite{DeRujula:2010:NotEndangered, DeRujula:2011:QEDconfronts}, scattering
experiments cannot completely exclude the existence of a ``thorn'' or a
``lump'' in the form factor $G_E(Q^2)$ at extremely low-$Q^2$
regime~\cite{Wu:2011} which could give rise to unexpectedly large
higher moments of the charge distribution.
Nevertheless, such a low-$Q^2$ behaviour is disfavoured 
by chiral perturbation theory ($\chi$PT) and vector meson
dominance (VMD) models~\cite{LorenzHammerMeissner:2012} which account
for the pion cloud in the low-energy regime.
It was demonstrated by Pineda that \rZt{} is about 2-3~fm$^3$ in the
leading chiral expansion
term~\cite{Pineda:2005:Lamb-shift:proton-radius-def, Pineda:2012}.
This sets tight constraints on the long tail of the proton charge
distribution.
Still it is important to further investigate the proton structure at
very low $Q^2$ with various techniques from $\chi$PT, to lattice QCD
and VMD.

\subsubsection{Radiative and higher-order corrections to the finite-size effect}

Radiative corrections to the finite-size contributions are listed in
Table~\ref{tab:fs}.
Using a perturbative approach Borie calculated the finite-size
correction to one-loop eVP ($-0.0110\langle r^2\rangle$~meV), the
finite-size correction to one-loop eVP-iteration ($-0.0165\langle
r^2\rangle$~meV) and the radiative-correction to two-photon
exchange~\cite{Eides:1997:two-photon} ($-0.00062\langle
r^2\rangle$~meV).
Similar results have been obtained by Pachucki~\cite{Pachucki:1999:ProtonMup}
and Karshenboim~\cite{Karshenboim:2012:relrecoil}.
Indelicato has also re-evaluated the main radiative corrections accounting for 
finite-size effects~\cite{Indelicato:2012}.
He computed the all-order finite-size corrections to the all-order
one-loop eVP (one-loop eVP + eVP iteration) and to the
K\"all$\acute{\mathrm{e}}$n-Sabry (including eVP iteration) term,
using a dipole charge distribution.
Moreover Indelicato and Mohr calculated the finite-size correction to
the muon self-energy perturbatively, and  confirmed their
result by using also an all-order finite-size
approach~\cite{Indelicato:Mohr:2012, Indelicato:2012}.
Finite-size corrections to higher-order radiative contributions are
negligible.

The total radiative corrections to finite-size in one-photon exchange (OPE)
is given by 
\begin{equation}
\label{eq:ope_rad}
\Delta E_\mathrm{OPE}^\mathrm{rad} =
  -0.0282\rr{} - 0.0002\rr{} + 0.0009\rr{} ~ \mathrm{meV} =
  -0.0275 \rr{}  ~ \mathrm{meV}.
\end{equation}

The higher-order finite-size correction given in Table~\ref{tab:fs}
for Borie is $\Delta E_\mathrm{finite\; size}^\mathrm{ho, \;
  Borie}=-0.000123$~meV~\cite{Borie:2012:LS_revisited} originating
from the terms in Eq.~(\ref{eq:friar}) not proportional to $\langle
r^2 \rangle$ and \rZt{}.
For Indelicato the higher-order finite-size correction $\Delta
E_\mathrm{finite\; size}^\mathrm{ho, \;
  Indelicato}=0.00001(10)$~meV~(Eq.~(114) in \cite{Indelicato:2012}) is the sum of
the terms in Eq.~(\ref{eq:fs-paul}) beyond $\rpC^2$ and $\rpC^3$ and
the radiative finite-size corrections beyond the $\langle r^2 \rangle$
terms for one-loop eVP, K\"all$\acute{\mathrm{e}}$n-Sabry and muon
self-energy.

\subsubsection{Summary of finite-size contributions}

From Table~\ref{tab:fs} we see that the finite-size contributions are
(OPE: one-photon exchange, TPE: two-photon exchange)
\begin{equation}
\label{eq:finite-size-numerical}
\begin{array}{lll}
\Delta E_\mathrm{finite\; size}^\mathrm{th}& =&
\Delta E_\mathrm{OPE} + 
\Delta E_\mathrm{OPE}^\mathrm{rad} +
\Delta E_\mathrm{TPE} +
\Delta E_\mathrm{TPE, el.}^\mathrm{rad} +
\Delta E_\mathrm{TPE, inel.}^\mathrm{rad} +
\Delta E_\mathrm{finite\; size}^\mathrm{ho} \\
                                        &=&-5.1994\, \rpC^2  -0.0275\,\rpC^2 + 0.0332(20)      -0.00062\,\rpC^2+0.00018 \\
                                        & &+ 0.00001(10)   ~   \textrm{meV}.\\
\end{array}
\end{equation}
Here,
$\Delta E_\mathrm{OPE}$ is given by Eq.~(\ref{eq:ope_main}),
$\Delta E_\mathrm{OPE}^\mathrm{rad}$ is from Eq.~(\ref{eq:ope_rad}), and
$\Delta E_\mathrm{TPE}$ is from Eq.~(\ref{eq:tpe}),
$\Delta E_\mathrm{TPE, el.}^\mathrm{rad} = -0.00062\,\rpC^2$ are radiative 
corrections to the elastic part of the TPE contribution calculated by 
Borie~\cite{Borie:2012:LS_revisited},
$\Delta E_\mathrm{TPE, inel.}^\mathrm{rad}$ is the sum of eVP (0.00019\,meV) and 
SE (-0.00001\,meV) corrections to the (inelastic) proton 
polarizability~\cite{Borie:2012:LS_revisited} 
(sum of \#26 and  \#27 in Table~\ref{tab:fs}), and 
$\Delta E_\mathrm{finite\; size}^\mathrm{ho}$ are higher-order finite size 
corrections~\cite{Indelicato:2012}. Each term corresoponds to one block in Table~\ref{tab:fs}.
In sum we obtain
\begin{equation}
\label{eq:finite-size-numerical_compressed}
\Delta E_\mathrm{finite\; size}^\mathrm{th}  =  -5.2275(10)\, \rpC^2 +0.0332(20)+ 0.0002(1) ~ \textrm{meV}.\\
\end{equation}
The uncertainty of the $\rpC^2$ coefficient accounts for the
difference between the all-order and perturbative approach and
uncertainties related to the proton charge
distribution.  Including the contributions summarized in
Table~\ref{tab:muonic-h-lamb} we obtain the total Lamb shift
prediction
\begin{equation}
\label{eq:lamb-shift}
\begin{array}{ll}
\Delta E_\mathrm{L}^\mathrm{th}   &= 206.0336(15) -5.2275(10)\, \rpC^2 +0.0332(20)     ~ \textrm{meV}\\[2mm]
                              &= 206.0668(25) -5.2275(10)\, \rpC^2   ~  \textrm{meV.}
\end{array}
\end{equation}
Note that the third Zemach moment or $\rpC^3$ contribution has now
been accounted for in a more appropriate quantum field framework by the
full TPE contribution $\Delta
E_\mathrm{TPE}$. Equation~(\ref{eq:lamb-shift}) can be compared with
the results in Ref.~\cite{Pohl:2010:Nature_mup1_suppl}
\begin{equation}
\label{eq:old-nature}
\begin{array}{ll}
\Delta E_\mathrm{L}^\mathrm{th}
 & = 206.0573(45)-5.2262\, \rpC^2 +0.0347\, \rpC^3 ~ \textrm{ meV}\\[2mm]
 & = 206.0779(45)-5.2262\, \rpC^2 ~ \textrm{ meV}
\end{array}
\end{equation}
and in Borie-v6 
\begin{equation}
\label{eq:borie-ls}
\begin{array}{ll}
\Delta E_\mathrm{L}^\mathrm{th} & =  206.0592(60)-5.2272\, \rpC^2 +0.0365(18)\, \rpC^3 ~ \textrm{meV}\\[2mm]
                            & =  206.0808(61)-5.2272\, \rpC^2~ \textrm{meV}\,.
\end{array}
\end{equation}
The difference in the constant terms between Eq.~(\ref{eq:lamb-shift})
and Eqs.~(\ref{eq:old-nature}) and (\ref{eq:borie-ls}) originates
mainly from item \#18 in Table~\ref{tab:fs} (0.013 meV) which was
double-counted in Ref.~\cite{Pohl:2010:Nature_mup1_suppl}.
Another double counting in
Ref.~\cite{Pohl:2010:Nature_mup1_suppl} was related to \#21 and \#12.
The uncertainty of the
proton structure independent term in Eq.~(\ref{eq:lamb-shift}) is
given mainly by the uncertainties of the radiative-recoil correction
\#24 and uncalculated higher-order terms.

\section{2S-HFS and the Zemach radius}
\label{sec:HFS_zemach}

The interaction between the bound particle and the magnetic field
induced by the magnetic moment of the nucleus gives rise to shifts and
splittings of the energy levels termed hyperfine effects.
In classical electrodynamics the interaction between the magnetic
moments \mbox{\boldmath{$\mu$}}$_p$, and \mbox{\boldmath{$\mu$}}$_\mu$
of proton and muon, respectively, is described by~\cite{Eides:2001:HydTheo}
\begin{equation}
\label{eq:magnetic-moments}
H_\mathrm{HFS}^\mathrm{classical}  =  -\dfrac{2}{3} \mbox{\boldmath{$\mu$}}_p \cdot  \mbox{\boldmath{$\mu$}}_\mu  \delta(\mbox{\boldmath{$r$}}) 
\end{equation}
where $\delta(\mbox{\boldmath{$r$}})$ is the delta-function in
coordinate space. A similar Hamiltonian can be derived in quantum
field theory from the one-photon exchange diagram. Using the Coulomb
wave function this gives rise in first order perturbation theory to an
energy shift for muonic hydrogen nS-states
of~\cite{Borie:2012:LS_revisited}
\begin{equation}
\label{eq:fermi-energy}
\begin{array}{lcl}
E_\mathrm{HFS}(F)& = &  \dfrac{4 (Z\alpha)^4 m_r^3}{3 n^3 m_\mu m_p}(1+\kappa)(1+a_\mu)\; \dfrac{1}{2}\Big[ F(F+1)-\dfrac{3}{2}\Big]\\[5mm]
                      & = & \Delta E_\mathrm{Fermi}\; \dfrac{1}{2} \Big[ F(F+1)-\dfrac{3}{2}\Big]
\end{array}
\end{equation}
where $\Delta
E_\mathrm{Fermi}=22.8320$~meV\cite{Borie:2012:LS_revisited} is the
Fermi energy, $m_p$ the proton mass, $F$ the total angular momentum,
$\kappa$ and $a_\mu$ the proton and muon anomalous magnetic moments,
respectively.
The $F=1$ state is shifted by $1/4\times 22.8320$~meV whereas the
$F=0$ state by $-3/4\times 22.8320$~meV (see
Fig.~\ref{fig:energy-levels}).
Equation~(\ref{eq:fermi-energy}) accounts for the sum of the terms
(h1) and (h4) in Table~\ref{tab:hfs}.
The Breit term (h2) corrects for relativistic and binding effects
accounted for in the Dirac-Coulomb wave function, but excluded in the
Schr\"odinger wave function.

Table~\ref{tab:hfs} also summarizes the corrections arising from QED,
recoil, nuclear structure, hadronic and weak interaction effects.
The structure-dependent corrections, scaling as the reduced mass of
the system, become large in \mup{} compared to hydrogen.
The largest correction is thus given by finite-size effect which, in
the non-relativistic limit, is given by the well known Zemach term
(h20)~\cite{Zemach:1956,Friar:1978:Annals}
\begin{equation}
\label{eq:HFS-finite-size}
\Delta E_\mathrm{Zemach}  = -\Delta E_\mathrm{Fermi}\cdot 2 (Z\alpha) m_r \;\rZ
\end{equation}
where \rZ{} is the Zemach radius defined as
\begin{equation}
\label{eq:rZ}
\rZ=  \int \mathrm{d}^3r \int \mathrm{d}^3r' \; r \,\rho_E (\mbox{\boldmath{$r$}})\rho_M (\mbox{\boldmath{$r$}}-\mbox{\boldmath{$r'$}}) 
\end{equation}
with $\rho_M $ and $\rho_E $ being the normalized proton magnetization
and charge distributions, respectively.
The convolution between charge and magnetization distribution in \rZ\ is a
consequence of the interaction of the proton spin distributed
spatially (given by the magnetic form factor) with the spatial
distribution of the muon spin which is described by the atomic muon
wave function. The latter is slightly affected, notably at the origin,
by the charge-finite-size effect and thus by $\rho_E$.
In a quantum field framework this contribution arises from two-photon
exchange processes. 
Similar to the situation for the
Lamb shift discussed above,  the intermediate virtual proton 
may be either ``on-shell'' or ``off-shell''.
Hence polarizability contributions need to be accounted for (h22).
This term has the largest uncertainty. It arises from the
uncertainty of the polarized structure functions $g_1$~\cite{Prok:2009, Anthony:2000} and $g_2$~\cite{Wesselmann:2007}
(measured in inelastic polarized electron-proton scattering) needed as
an input to calculate this contribution.
For the HFS (in contrast to the Lamb shift) no subtraction term is required
for the calculation of the two-photon exchange diagrams via Compton scattering
and dispersion analysis~\cite{Carlson:2008:p_struct_HFS}.

The leading recoil correction to the HFS (h23) is generated by the
same two-photon exchange diagram and is of order $(Z\alpha) (m/M)
\tilde{E}_\mathrm{Fermi}$ where $\tilde{E}_\mathrm{Fermi}$ is the
Fermi energy without contribution of the muon anomalous magnetic
moment~\cite{Eides:2001:HydTheo}.

Radiative corrections which are not accounted for by
the anomalous magnetic moment are listed separately in Tab.~\ref{tab:hfs}.
The largest radiative correction is related to the distortion of the
wave functions caused by the Uehling potential, items (h5),
(h7), (h11) and (h25).
Other corrections account for modifications of the magnetic
interaction caused by the eVP in one- and two-photon exchange (h8),
(h9) and (h10), and vertex corrections caused mainly by the muon
self-energy (h13) and (h14).

The main HFS contributions have been confirmed and refined by
Indelicato~\cite{Indelicato:2012} by numerical integration of the
hyperfine Hamiltonian with Bohr-Weisskopf (magnetization distribution)
correction using Dirac wave functions. The latter have been calculated
for Coulomb finite-size and Uehling potentials.  All-order
finite-size, relativistic and eVP effects are thus included in the
wave function.
This calculation is performed for various \rpC{} and \rZ{}, assuming
exponential charge and magnetization distributions.
From these calculations the terms (h3), (h6), (h20), (h21) and (h25)
are obtained, showing good agreement with the perturbative results.
It is interesting to note that the HFS shows a small dependence on
$\rpC^2$ given by the term (h21).
The small constant terms in (h21) and (h25) account for the sum of
higher order terms of a polynomial expansion in \rpC{} and \rZ{}
($\rZ^2$, $\rpC \rZ$, $\rpC^2 \rZ$, $\rpC \rZ^2$, $\rpC^2 \rZ^2$ etc.)
of the numerical results obtained in~\cite{Indelicato:2012}.

\begin{table*}[t!]
\renewcommand{\baselinestretch}{1.1}
\renewcommand{\arraystretch}{1.5}
  \caption{All known contributions to the 2S-HFS in $\mu$p from
    different authors, and the one we selected (usually the all-order
    calculations which we consider more complete).  Values are in meV,
    radii in fm.  SE: self-energy, VP: vacuum polarization,  Rel: relativistic, 
    RC: recoil correction, PT: perturbation theory, p: proton, int: interaction, AMM: anomalous magnetic moment.}
  \label{tab:hfs}
\setlength\tabcolsep{1mm}
{
  \fontsize{7pt}{7pt}\selectfont
  \begin{tabular}{r|l| r@{.}l  r@{.}l r@{.}l r@{.}l l}
  \hline
  \hline
 & Contribution                                                                    & \lft{2}{Martynenko}                 & \lft{2}{Borie-v6} & \lft{2}{Indelicato} & \multicolumn{2}{l}{Our choice}   & Ref. \\
 &                                                                                 & \lft{2}{\cite{Martynenko:2005:mupHFS2S}}& \lft{2}{\cite{Borie:2012:LS_revisited}} &\lft{2}{ \cite{Indelicato:2012}}\\
\hline 
 h1 & Fermi energy,  $(Z\alpha)^4$                                                 &          22&8054                             &     22&8054                        & \cnt{2}{}     & \cnt{2}{}            &  \\
 h2 & Breit corr., $(Z\alpha)^6$                                              &          0&0026                              &      0&00258                       & \cnt{2}{}     &\cnt{2}{}             & \\
 h3 & Dirac energy (+ Breit corr. in all-order)                             &        \cnt{2}{}                             &     \cnt{2}{}                      & 22&807995     &     22&807995        & Eq.~(107) in \cite{Indelicato:2012}\\[2mm]
  
  h4 & $\mu$ AMM corr., $\alpha(Z \alpha)^4$, $\alpha²(Z \alpha)^4$&           0&0266                            &      0&02659                        & \cnt{2}{}    &       0&02659         & \\
\hline
  h5 & eVP in 2nd-order PT, $\alpha (Z\alpha)^5$ ($\epsilon_\mathrm{VP2}$)&           0&0746                           &      0&07443                        & \cnt{2}{}    & \cnt{2}{}              &\\
  h6 & All-order eVP corr.                                                           &         \cnt{2}{}                          &     \cnt{2}{}                       &  0&07437    &       0&07437       &  Eq.~(109) in \cite{Indelicato:2012}\\
  h7 & Two-loop corr. to Fermi-energy ($\epsilon_\mathrm{VP2}$)                   &         \cnt{2}{}                         &      0&00056                        & \cnt{2}{}    &       0&00056         &\\
  \hline
  h8 & One-loop eVP in $1\gamma$ int., $\alpha (Z\alpha)^4$ ($\epsilon_\mathrm{VP1}$)&      0&0482                           &      0&04818                       & \cnt{2}{}    &       0&04818         & \\
  h9 & Two-loop eVP in  $1\gamma$ int., $\alpha^2 (Z\alpha)^4$ ($\epsilon_\mathrm{VP1}$)&  0&0003                             &      0&00037                      & \cnt{2}{}    &       0&00037         & \\
 h10 & Further two-loop eVP corr.                                              &          \cnt{2}{}                         &      0&00037                       & \cnt{2}{}    &       0&00037         &~\cite{Karshenboim:2008:JETP_Hyperfine, Karshenboim:2008:JETP_Hyperfine_erratum} \\ 
 \hline
 h11 & $\mu$VP (similar to  $\epsilon_\mathrm{VP2}$)                                   &           \cnt{2}{}                         &      0&00091                     & \cnt{2}{}    &       0&00091         & \\
 h12 & $\mu$VP (similar to  $\epsilon_\mathrm{VP1}$)                                   &           0&0004                            &     \multicolumn{2}{l}{(incl. in h13)}  & \cnt{2}{} & \multicolumn{2}{l}{(incl. in h13) }  &    \\
 \hline
 h13 & Vertex, $\alpha (Z\alpha)^5$                                                  &          \cnt{2}{}                          &      -0&00311                     & \cnt{2}{}  &       -0&00311         & \footnote{Includes a correction $\alpha (Z\alpha)^5$ due to $\mu$VP. }\\
 h14 & Higher order corr. of (h13), (part with $\ln(\alpha)$)                     &          \cnt{2}{}                          &      -0&00017 􀀀                   & \cnt{2}{}  &       -0&00017         &~\cite{Brodsky:1966:radiative:hyperfine} \\[2mm]
h15 & $\mu$ SE  with p structure,  $\alpha (Z\alpha)^5$               &           0&0010                             &     \cnt{2}{}                    & \cnt{2}{}  & \cnt{2}{}                & \\
h16 & Vertex corr. with proton structure,  $\alpha (Z\alpha)^5$                    &           -0&0018                            &    \cnt{2}{}                     & \cnt{2}{}  & \cnt{2}{}               & \\
h17 & ``Jellyfish'' corr. with p structure,  $\alpha (Z\alpha)^5$                   &            0&0005                            &     \cnt{2}{}                    & \cnt{2}{}  & \cnt{2}{}               & \\
\hline
h18 & Hadron VP,  $\alpha^6$                                                        &           0&0005(1)                          &      0&00060(10)                 & \cnt{2}{}  &       0&00060(10)          & \\
h19 & Weak interaction contribution                                                 &           0&0003                             &      0&00027                     & \cnt{2}{}  &        0&00027             & \cite{Eides:2012:Weak}\\
\hline
h20 & Finite-size  (Zemach) corr. to $\Delta E_\mathrm{Fermi}$, $(Z\alpha)^5$􀀀       &          -0&1518\footnote{Calculated using the Simon et al. form factor.} 􀀀&\lft{2}{-0.16037\,\rZ{} \;\;} & \multicolumn{2}{l}{\,-0.16034\,\rZ{}\;\;} & \multicolumn{2}{l}{\,-0.16034\,\rZ{}\;\;} & Eq.~(107) in \cite{Indelicato:2012} \\[2mm]
h21 & Higher-order finite-size corr. to $\Delta E_\mathrm{Fermi}$                    &\cnt{2}{ }                                    &\cnt{2}{ }                                     & \multicolumn{2}{l}{\,-0.0022\,$\rpC{}^2$} & \multicolumn{2}{l}{\,-0.0022\,$\rpC{}^2$} & Eq.~(107) in \cite{Indelicato:2012}\\[0mm]
    &                                                                              &   \cnt{2}{ }                                 &\cnt{2}{ }                                     & \multicolumn{2}{l}{\,+0.0009}               & \multicolumn{2}{l}{\,+0.0009}  & \\[2mm]   

h22 & Proton polarizability,  $(Z\alpha)^5$,􀀀 $\Delta E_\mathrm{HFS}^\mathrm{pol}$    & 0&0105(18)                                   &      0&0080(26)                    & \cnt{2}{}            &    0&00801(260) & ~\cite{CarlsonNazaryan:2011,Cherednikova:2002}\\
\hline 
h23 & Recoil corr.                                                            &\multicolumn{2}{l}{(incl. in h20)}           &      0&02123                      & \cnt{2}{}              &    0&02123           &~\cite{Carlson:2008:p_struct_HFS}\\      
\hline
h24 & eVP + proton structure corr.,  $\alpha^6$                                    &            -0&0026                            & \cnt{2}{}                      & \cnt{2}{}                 & \cnt{2}{}                & \\
h25 & eVP corr. to finite-size (similar to  $\epsilon_\mathrm{VP2}$)                 &\cnt{2}{}                                       & -0&00114                        & \multicolumn{2}{l}{\,-0.0018\,\rZ{}}               & \multicolumn{2}{l}{\,-0.0018\,\rZ{}} & Eq.~(109) in \cite{Indelicato:2012}\\[0mm]
    &                                                                               &\cnt{2}{}                                       &\cnt{2}{}                      & \multicolumn{2}{l}{\,-0.0001}    & \multicolumn{2}{l}{\,-0.0001} \\[2mm]      
h26    & eVP corr. to finite-size (similar to  $\epsilon_\mathrm{VP1}$)              &\cnt{2}{}                                       & -0&00114                      &\cnt{2}{}                  & -0&00114(20)    \\[2mm]

 h27 & Proton structure corr., $\alpha (Z\alpha)^5$                                 & -0&0017 &  \cnt{2}{}                                   &  \lft{2}{}        &  \cnt{2}{} \\
 h28 & Rel. + radiative RC with p AMM,   $\alpha^6$                          &           0&0018                                 & \cnt{2}{}                                      & \cnt{2}{}         &  \cnt{2}{} \\
 \hline                                                                                                                            
 & Sum   & 22&8148(20)~\footnote{The uncertainty is 0.0078~meV if the uncertainty of the Zemach term (h20) is included (see Table II of ~\cite{Martynenko:2005:mupHFS2S}).}          & 22&9839(26)     &\cnt{2}{}           & 22&9858(26) & \\
 &                                                                                        & \lft{2}{} & \multicolumn{2}{l}{\;-0.1604\,\rZ{}} &\cnt{2}{}  &\multicolumn{3}{l}{\,-0.1621(10)\,\rZ{}\,-\,0.0022(5)\,$\rpC^2$}\\
 \hline
 & Sum  with $\rpC=0.841$~fm, $\rZ=1.045$ fm~\cite{Distler:2011:Zemach}                 & 22&8148 meV          & 22&8163  meV   &\cnt{2}{}           & 22&8149 meV & \\
\hline
\hline
  \end{tabular}
}
\end{table*}

%
The sum of all contributions in Table~\ref{tab:hfs} is
\begin{equation}
\label{eq:HFS}
\Delta E_\mathrm{HFS}^\mathrm{th}  =  22.9763(15) - 0.1621(10)\rZ + \Delta E^\mathrm{pol}_\mathrm{HFS} ~ \textrm{meV}
\end{equation}
where \rZ{} is in fm.
Using the value $\Delta
E^\mathrm{pol}_\mathrm{HFS}=0.0080(26)$~meV~\cite{CarlsonNazaryan:2011}
results in
\begin{equation}
\label{eq:HFS-1}
\Delta E_\mathrm{HFS}^\mathrm{th}=    22.9843(30)- 0.1621(10)\rZ ~ \textrm{meV}.
\end{equation}
The uncertainty of the first term in Eq.~(\ref{eq:HFS}) considers
differences between results from various authors and uncalculated
higher-order terms.

\section{Conclusions}

We have presented an update of the theoretical predictions 
for the 2S Lamb shift in muonic hydrogen (see Eq.~(\ref{eq:lamb-shift}))
\begin{equation}
\begin{array}{ll}
\Delta E_\mathrm{L}^\mathrm{th}   &= 206.0336(15) -5.2275(10)\, \rpC^2 +0.0332(20)     ~ \textrm{meV}\\[2mm]
                              &= 206.0668(25) -5.2275(10)\, \rpC^2     ~ \textrm{meV}.
\end{array}
\end{equation}
Similarly, the 2S hyperfine splitting (HFS) in muonic hydrogen is (see Eq.~(\ref{eq:HFS})) 
\begin{equation}
\begin{array}{ll}
 \Delta E_\mathrm{HFS}^\mathrm{th} & =  22.9763(15) - 0.1621(10)\rZ + 0.0080(26) ~ \textrm{meV}\\[2mm]
                                & =    22.9843(30)- 0.1621(10)\rZ ~ \textrm{meV}.
\end{array}
\end{equation}
Double-counting of a few higher-order terms in previous compilations
has been eliminated.
No large error and no missing contributions beyond 0.001~meV have been found.
Finite-size and one-loop eVP has recently been studied also by 
numerical integration of the Dirac equation including the finite-size
Coulomb potential and Uehling potential~\cite{Indelicato:2012}, confirming the 
perturbative results, for both the Lamb shift and the HFS.
The uncertainty arising from the two-photon exchange is still debated
but large contributions seem unlikely.
The total unertainty of the Lamb shift theory has been reduced by a factor of
2 since the summary~\cite{Pohl:2010:Nature_mup1_suppl}.

The 0.3 meV ($7\sigma$) discrepancy between the proton rms charge radii \rpC{}
from muonic hydrogen~\cite{Pohl:2010:Nature_mup1,Antognini:2012:Nature_mup2}
and CODATA-2010~\cite{Mohr:2012:CODATA10} persists.
The ``proton radius puzzle'' remains.


\section{Acknowledgments}
We thank Michael Birse, Edith Borie, Stan Brodsky, Carl Carlson,
Michael Distler, Michael Eides, Jim Friar, Richard Hill, Ulrich
Jentschura, Savely Karshenboim, Klaus Kirch, Gerald A.\ Miller, Peter
Mohr, Krzysztof Pachucki, Gil Paz, Antonio Pineda, Roland Rosenfelder,
Ingo Sick, Adrian Signer for valuable comments.  We acknowledge support from the
Swiss National Science Foundation project 200021L--138175/1.
R.P.\ acknowledges support from the European Research Council (ERC) through
Starting Grant \#279765.

\bibliography{refs}
\end{document}